%% file: puffin_ann_arxiv.tex
\documentclass[11pt,a4paper]{article}

\usepackage[utf8]{inputenc}
\usepackage[T1]{fontenc}
\usepackage[margin=1in]{geometry}
\usepackage{amsmath,amssymb}
\usepackage{graphicx}
\usepackage{booktabs}
\usepackage{listings}
\usepackage{xcolor}
\usepackage{tikz}
\usetikzlibrary{positioning,arrows.meta,fit,calc,shapes.geometric,shapes.multipart}
\usepackage{hyperref}
\usepackage{url}
\usepackage{authblk}
\usepackage{caption}

\hypersetup{
  colorlinks=true,
  linkcolor=black,
  citecolor=black,
  urlcolor=blue
}

\definecolor{codegray}{rgb}{0.95,0.95,0.95}
\title{Puffin-Backed Vector Indexes:\\
Attaching Approximate Nearest Neighbor Indexes to\\
Apache Iceberg Snapshots for Compute-Disaggregated Query Engines}

\author[1]{Artur Borycki}
\affil[1]{Teradata Advanced Research \\ \texttt{artur.borycki@teradata.com}}

\date{April 2026}

\begin{document}

\maketitle

\begin{abstract}
We describe a design pattern and concrete implementation for embedding distributed approximate nearest neighbor (ANN) indexes inside the Apache Iceberg table format, using the Puffin sidecar file as the storage container and the snapshot summary as the binding mechanism. Modern analytical query engines increasingly adopt a compute-disaggregated architecture: executors are stateless, scale elastically, and read all data from object storage. Adding vector similarity search to such an engine traditionally requires a dedicated index storage layer with its own lifecycle, consistency model, and operational surface --- breaking the disaggregation invariant. We show that the Puffin format, originally introduced for table-level statistics and deletion vectors, is sufficient to carry full Vamana/DiskANN graphs at billion-vector scale, and that linking these blobs through the existing \texttt{statistics-file} snapshot summary property reduces ANN index management to standard Iceberg snapshot operations. We present (i) a binary layout for sharded graph indexes inside Puffin, (ii) a coordinator--executor protocol for distributed index build, probe, and incremental refresh, (iii) the integration into the existing optimistic-concurrency commit path of an Iceberg REST catalog, and (iv) a tiered probe strategy that places small centroid indexes on the coordinator and large DiskANN graphs on executor SSDs. The pattern inherits atomicity, time travel, multi-engine readability, and orphan-file garbage collection from the table format at zero implementation cost. We discuss the recall/latency trade-offs introduced by the independent-shard design and quantify projected query and build performance for tables up to $10^9$ vectors. Our implementation extends FlockDB, a distributed MPP engine built on DuckDB, and requires approximately 8{,}000 lines of new code, the majority of which is the graph builder itself; the integration with the table format is a few hundred lines.
\end{abstract}

\vspace{1em}

\section{Introduction}

The rise of embedding-based retrieval, retrieval-augmented generation, and large-scale semantic search has made vector similarity search a first-class operation in analytical systems~\cite{douze2024faiss,wang2021milvus}. The standard architectural response has been the dedicated vector database: a stateful service that owns its index storage, its query path, and its consistency semantics. This works well for online serving workloads, but it sits awkwardly alongside the modern analytical data lake.

Two architectural trends are in tension. First, analytical engines are converging on a \emph{compute-disaggregated} design~\cite{armbrust2021lakehouse,vuppalapati2020building}: data is stored as immutable files (typically Parquet~\cite{parquet}) in object storage, organized by an open table format (Apache Iceberg~\cite{iceberg-spec}, Delta Lake~\cite{armbrust2020delta}, Apache Hudi~\cite{hudi}), and queried by stateless executor processes that can be added or removed without data movement. Second, vector indexes for billion-scale corpora are themselves multi-tens-to-hundreds of gigabytes~\cite{subramanya2019diskann,jayaram2019spann} and have non-trivial build cost and refresh semantics. Adding such indexes to a disaggregated engine without breaking its design tenets is non-obvious.

The conventional integration paths --- running a separate vector database, maintaining a private indexed filesystem on executors, or rebuilding the index on every query --- each violate at least one of the following properties that a lakehouse engine wants to preserve:

\begin{itemize}
  \item \textbf{Stateless executors.} Executors must be safe to terminate and replace at any moment.
  \item \textbf{Single source of truth.} Derived state (indexes, statistics) must remain consistent with the data version it was computed against.
  \item \textbf{Multi-engine interoperability.} The table format permits multiple engines to read the same data; derived artifacts should not break that.
  \item \textbf{No new lifecycle.} Index creation, refresh, and deletion should reuse the table format's transactional commit, time-travel, and garbage-collection mechanisms.
\end{itemize}

We observe that the Apache Iceberg specification already includes a mechanism that satisfies all four of these properties for arbitrary derived data: the \emph{Puffin file format}~\cite{iceberg-puffin-spec}. Puffin was introduced for table statistics (Theta sketches for distinct-value estimation, bloom filters for partition pruning) and deletion vectors, but the format is intentionally extensible: each blob carries an opaque \texttt{type} string, an Iceberg field-ID set, an arbitrary property map, and a payload of bytes. A Puffin file is bound to a snapshot via a single property in the snapshot summary (\texttt{statistics-file}).

This paper argues that Puffin is the correct storage container for ANN indexes in disaggregated lakehouse engines, and describes a complete design for using it that way. Our contributions are:

\begin{enumerate}
  \item \textbf{A design pattern} (Section~\ref{sec:design}) that maps Vamana/DiskANN~\cite{subramanya2019diskann} graph indexes onto Puffin blobs, with one blob per executor shard plus a routing blob, achieving build parallelism that scales linearly with executors while inheriting all consistency guarantees of the underlying table format.
  \item \textbf{A byte-level format specification} (Section~\ref{sec:format}) for two blob types: a small file-level centroid index suitable for coordinator-local probing, and a sharded Vamana graph layout suitable for executor-local SSD-resident probing.
  \item \textbf{A distributed build, probe, and refresh protocol} (Sections~\ref{sec:build}--\ref{sec:refresh}) that reuses the engine's existing INSERT-SELECT write path, its cache-aware scheduler, and its snapshot diffing mechanism, with only the graph operations themselves being genuinely new code.
  \item \textbf{An incremental refresh algorithm} (Section~\ref{sec:refresh}) that exploits Iceberg's manifest-level diff to identify only the changed Parquet files, applies Vamana greedy insert for added vectors and lazy tombstoning for removed ones, and commits the updated Puffin file as a metadata-only snapshot operation.
  \item \textbf{Quantitative analysis} (Section~\ref{sec:eval}) of the projected query latency, build time, and storage cost for a representative billion-vector workload, along with an honest discussion of the recall trade-off introduced by the independent-shard design (Section~\ref{sec:limitations}).
\end{enumerate}

We implement this design in FlockDB, a distributed MPP analytical engine built on DuckDB~\cite{raasveldt2019duckdb} and Apache Iceberg, deployed against an AIStor object store. The implementation comprises approximately 8{,}000 lines of new code; we report on the structure of the implementation rather than empirical benchmarks, which remain ongoing.

We frame this work as a systems design paper. The novelty is not in the ANN algorithm --- we use Vamana with product quantization~\cite{jegou2011pq} essentially unchanged --- but in the storage and lifecycle integration. To our knowledge no prior published system has placed a billion-scale ANN graph inside an Iceberg-format sidecar and used the table format's commit path to manage its lifecycle.

\section{Background and Related Work}
\label{sec:background}

\subsection{Open Table Formats and Puffin}

Apache Iceberg~\cite{iceberg-spec} is an open table format that layers transactional semantics, schema evolution, partition evolution, and time travel over immutable Parquet files in object storage. A table is described by a chain of immutable \emph{snapshots}, each referencing a manifest list, which in turn references manifest files describing the data files live at that snapshot. Snapshot commit is atomic at the REST catalog layer using optimistic concurrency control.

The Puffin format~\cite{iceberg-puffin-spec} is a flat binary container introduced to hold table-level derived data that does not fit naturally into manifests. The file structure is straightforward: a four-byte magic number (\texttt{PFA1}), a concatenated sequence of opaque blob payloads, a UTF-8 JSON footer describing each blob (offset, length, type, field-ID set, compression codec, properties), a footer length, four flag bytes, and a trailing magic number. Each blob is independently compressible (zstd, lz4) and individually addressable via byte-range request, which matters for very large files where downloading the full content is impractical.

The two standardized blob types in the Iceberg ecosystem today are \texttt{apache-datasketches-theta-v1} for distinct-value estimation and \texttt{deletion-vector-v1} for row-level deletion bitmaps. The specification explicitly permits arbitrary additional types and properties, and several engines have begun using this extension point for engine-specific statistics. We are not aware of a published use of Puffin for ANN indexes prior to this work.

A Puffin file is bound to a snapshot through the \texttt{statistics-file} summary property: a free-form key-value map serialized as part of the snapshot. Any engine reading the snapshot can locate the Puffin file, inspect the footer, and selectively read blobs of types it understands; unknown blob types are ignored without error. This property forms the basis of our integration.

\subsection{Approximate Nearest Neighbor Search}

The literature on ANN is extensive; we summarize only the design points that matter for our integration. The three dominant families of indexes at billion-vector scale are inverted file with product quantization (IVF-PQ)~\cite{jegou2011pq}, hierarchical small-world graphs (HNSW)~\cite{malkov2018hnsw}, and Vamana graphs as used in DiskANN~\cite{subramanya2019diskann}. We focus on Vamana because it has two properties that fit a disaggregated engine particularly well.

First, Vamana graphs are designed for SSD-resident operation. The graph adjacency is memory-mapped from local disk; only the visited nodes (typically a few hundred per query) are paged into RAM. Combined with product quantization for the in-memory distance approximation, a 250M-vector shard fits comfortably in $\sim$4\,GB of RAM and $\sim$60\,GB of SSD --- well within the resource envelope of a modern analytical executor node. Second, Vamana supports a cheap incremental insert (greedy beam search to find candidate neighbors, robust-prune to enforce the degree bound, bidirectional edge insertion). This makes the index amenable to incremental refresh against new data files, which is essential for any lakehouse use case where the table is continuously appended.

A long line of work has explored placing ANN indexes alongside other data systems. Milvus~\cite{wang2021milvus} and Vespa~\cite{vespa} are purpose-built vector databases. PASE~\cite{yang2020pase} and pgvector~\cite{pgvector} embed ANN indexes inside PostgreSQL as extensions. SPANN~\cite{jayaram2019spann} demonstrates a hybrid memory/SSD layout. The work closest to ours in architectural style is AnalyticDB-V~\cite{wei2020analyticdb}, which extends a distributed analytical engine with vector indexes; however, AnalyticDB-V manages its own index storage rather than embedding the index in an open table format. We are not aware of prior work that uses the table format's own derived-data slot to carry ANN indexes.

\subsection{Compute-Disaggregated Query Engines}

The compute-disaggregated architecture, popularized by Snowflake~\cite{dageville2016snowflake}, separates a small persistent metadata service from a fleet of stateless compute nodes that mount object storage as their primary store. The lakehouse generalization~\cite{armbrust2021lakehouse} replaces the closed metadata service with an open table format. FlockDB, the host engine for our implementation, follows this design: a coordinator process holds query plans and the catalog client; executors are stateless processes with local SSDs used only as a cache. All durable state lives in object storage and in the table format catalog.

In such a system, any derived structure must be either (a) recomputable from the underlying data, or (b) committed into the table format itself. Local-disk-only state is incompatible with elastic scaling, because losing a node loses the work. Our design takes path (b): the ANN index is durably stored in Puffin files in object storage, and executor SSDs serve only as a cache for active shards.

\section{Design}
\label{sec:design}

\subsection{Architectural Overview}

\begin{figure}[h]
\centering
\resizebox{0.95\textwidth}{!}{%
\input{figures/tikz_overview.tex}%
}
\caption{System architecture. Compute (client, coordinator, executors) is stateless and elastic; storage (S3 plus REST catalog) is durable. A Puffin file lives in the same object store as Parquet data and is referenced by the snapshot summary's \texttt{statistics-file} property.}
\label{fig:overview}
\end{figure}
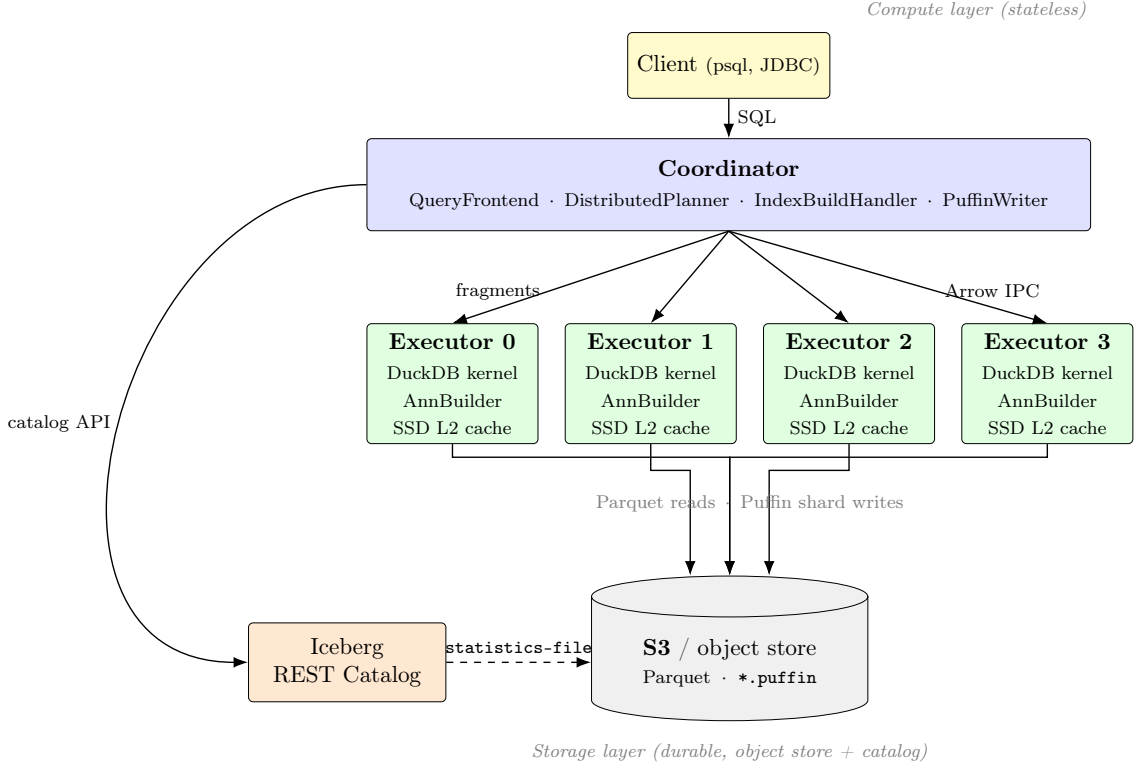

The system has three roles. A coordinator receives SQL queries via a PostgreSQL-compatible wire protocol, parses them, resolves table metadata through the Iceberg REST catalog, plans a distributed query, and dispatches work to executors. Executors are uniform processes that hold local SSD caches keyed by \texttt{(object\_path, credential\_fingerprint)} and a smaller in-memory L1 cache. Executors stream results back as Arrow IPC. The Iceberg REST catalog is the single source of truth for table state. Figure~\ref{fig:overview} sketches the components.

We add three components: an \texttt{IndexBuildHandler} on the coordinator that orchestrates index lifecycle (CREATE / REFRESH / DROP INDEX), an \texttt{AnnBuilder} on each executor that constructs and probes graph shards, and a \texttt{PuffinWriter} module that mirrors the existing read path to produce Puffin files in the required binary format. All other infrastructure --- distributed fragment dispatch, cache-aware scheduling, snapshot commit via REST catalog, snapshot diffing for incremental refresh --- is reused from the existing engine.

\subsection{Design Principles}
\label{sec:principles}

The design follows five principles.

\textbf{Principle 1: One blob per executor shard.} Rather than building a single global Vamana graph and partitioning it, each executor builds an independent local graph over its assigned subset of Parquet data files. The resulting graphs are stored as independent blobs in a single Puffin file. This trades a small recall loss (a global graph would find slightly better neighbors than $N$ independent graphs, recovered partially by an exact rerank phase) for two large architectural wins: build time scales linearly with executor count, and probe parallelism falls out of the existing fan-out/merge machinery the engine already uses for top-$K$ queries.

\textbf{Principle 2: Snapshot-bound index lifecycle.} Every Puffin file is bound to exactly one snapshot via the \texttt{statistics-file} summary property. Index versions are snapshot versions. Time travel queries (\texttt{AS OF}) automatically read the matching index version. Concurrent index builds are arbitrated by the same optimistic-concurrency mechanism that arbitrates concurrent data writes.

\textbf{Principle 3: Tiered probe placement.} Not every ANN index needs to be sharded across executors. A file-level centroid index for a 10{,}000-file table is around 30\,MB at 768 dimensions, and the probe (10{,}000 distance computations) takes under a millisecond on a single thread. We probe such small indexes on the coordinator directly, replacing the full file list with a pruned subset before dispatch. Only indexes that exceed a configurable size threshold (default 100\,MB) are sharded across executors.

\textbf{Principle 4: Reuse the write path.} An ANN index build is, structurally, an INSERT-SELECT that happens to produce a Puffin file instead of Parquet files. We dispatch index build fragments through the same \texttt{PlanFragment}/\texttt{TaskRunner} infrastructure that handles ordinary writes, with a new \texttt{IndexBuildTaskInfo} struct riding alongside the existing \texttt{WriteTaskInfo}. The Iceberg commit that finalizes the build is the same \texttt{CommitToIceberg} call, with one additional summary property.

\textbf{Principle 5: Incremental refresh via manifest diff.} The expensive operation in any ANN system is the rebuild. Iceberg already provides the primitive we need: given two snapshot IDs, the manifest diff returns precisely the data files added, removed, and unchanged between them. We refresh only the affected shards, using Vamana's native incremental insert for added vectors and a tombstone bitmap for removed ones.

\subsection{Tiered Probe Placement}
\label{sec:tiered}

Different ANN algorithms have very different size profiles. Table~\ref{tab:index-sizes} estimates index sizes for a one-billion-vector table at 768 dimensions.

\begin{table}[h]
\centering
\caption{Index size at $10^9$ vectors, 768 dimensions. PQ parameters: $m=48$, $\textit{nbits}=8$.}
\label{tab:index-sizes}
\small
\begin{tabular}{lrl}
\toprule
Index type & Size & Probe location \\
\midrule
File-level centroids ($10^4$ files) & $\sim$30\,MB & Coordinator \\
IVF-PQ ($\textit{nlist}=4096$) & $\sim$16\,GB & Borderline \\
HNSW ($M=16$) & $\sim$60\,GB & Executor (RAM) \\
DiskANN/Vamana ($R=64$) & $\sim$250\,GB & Executor (SSD) \\
\bottomrule
\end{tabular}
\end{table}

The coordinator probes any index smaller than \texttt{coordinator\_ann\_probe\_threshold\_mb} (default 100\,MB), which in practice means only file-level centroid indexes. Larger indexes are sharded across executors and probed with a three-stage distributed plan (Section~\ref{sec:probe}). This threshold is the cleanest way we found to balance two pressures: coordinator-side probing avoids an extra distributed stage and shaves off latency, but only the coordinator's RAM bounds it; executor-side probing scales horizontally and uses SSD, but pays a stage barrier of a few milliseconds.

\section{Puffin Format for ANN}
\label{sec:format}

We define two new Puffin blob types: \texttt{flockdb-ann-centroid-v1} for file-level centroid indexes and \texttt{flockdb-ann-index-v1} for Vamana graph shards. A third type, \texttt{flockdb-ann-routing-v1}, carries the metadata that ties the shards together.

\subsection{The Centroid Index Blob}

The centroid blob is a single self-contained binary structure: a 32-byte header (4-byte magic \texttt{ANNI}, version, dimensions, entry count, file count, metric code, entry size, file-paths section offset), followed by $N$ fixed-size entries each containing a $D$-dimensional float32 centroid, a 32-bit file index, and a 32-bit float \texttt{max\_distance} (the largest distance from the centroid to any vector in the represented file). After the entries comes a length-prefixed UTF-8 file paths table. The \texttt{max\_distance} field enables exact pruning for threshold queries: any file whose \texttt{centroid\_distance} minus its \texttt{max\_distance} exceeds the user's threshold can be eliminated safely.

For 10{,}000 files at 768 dimensions, the entry section is $10^4 \times (768 \times 4 + 8) = 30.8\,\text{MB}$. With zstd compression on the entire blob, this typically drops to 8--15\,MB depending on data distribution. The blob's \texttt{properties} map carries dimensions, metric, entry count, and the snapshot ID the centroids were computed against; the \texttt{fields} array contains the single Iceberg field ID of the indexed vector column.

\subsection{The Sharded Vamana Layout}

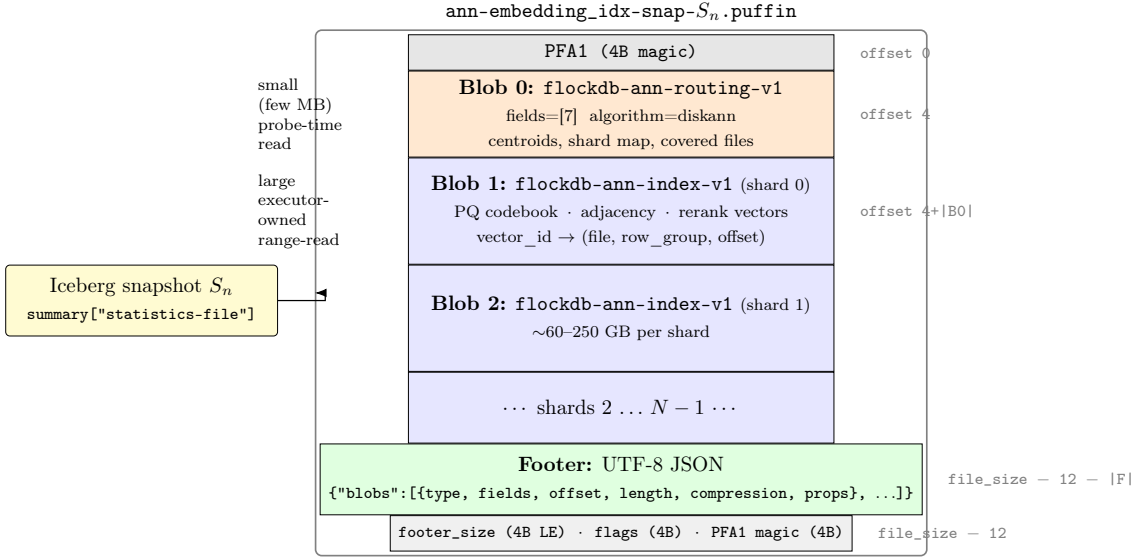
\begin{figure}[h]
\centering
\resizebox{0.95\textwidth}{!}{%
\input{figures/tikz_puffin_layout.tex}%
}
\caption{Puffin file layout for a sharded Vamana index. The Iceberg snapshot binds to the file via a single \texttt{statistics-file} summary property. Blob~0 (routing) is small and read on every probe to discover shard ownership; the shard blobs are large and downloaded only by their assigned executor via HTTP range request driven by the byte offsets in the footer JSON.}
\label{fig:puffin-layout}
\end{figure}

A sharded Vamana index occupies one Puffin file containing $N+1$ blobs: one \texttt{flockdb-ann-routing-v1} blob and $N$ \texttt{flockdb-ann-index-v1} blobs (one per executor shard). Figure~\ref{fig:puffin-layout} sketches the layout.

The routing blob is small (a few megabytes) and is read by every probe; the shard blobs are large (tens of gigabytes each) and are downloaded only by the executor that owns the shard. Puffin's footer-based random access is critical here: a coordinator can issue an HTTP range request for just the footer (typically a few kilobytes), parse it, and identify the byte offsets of each blob without downloading the full file. Executors then issue separate range requests for their assigned shard blobs.

\subsection{Shard Blob Internals}

Each shard blob contains, in order: a fixed header with magic \texttt{DANN}, version, dimensions, vector count, graph degree $R$, beam width $L$, medoid identifier, and PQ parameters; the PQ codebook ($K \times m$ centroids of $D/m$ dimensions each); an adjacency offset table ($N+1$ uint64 offsets); the adjacency lists themselves (zstd-compressed, with per-node degree and neighbor IDs as varints); the full float32 vectors used for exact rerank of the top candidates; and finally a vector-ID-to-location map providing \texttt{(file\_path, row\_group\_id, row\_offset)} tuples for each indexed vector. The map is stored in delta-encoded sorted order to keep its size manageable.

The full-vector section is the dominant cost ($N \times D \times 4$ bytes); for 250M vectors at 768 dimensions, this is approximately 760\,GB per shard. In practice we make this optional, with a configurable retention policy: if the host engine can re-fetch the vector from Parquet during the rerank phase, the shard blob omits this section entirely, dropping shard size from $\sim$760\,GB to $\sim$60\,GB. The trade-off is an extra round trip per query for vector retrieval during rerank.

\section{Distributed Build}
\label{sec:build}

Index construction follows a three-stage distributed plan.

\textbf{Stage 0: Sampling and centroid training (coordinator).} The coordinator samples approximately 1\% of vectors by reading a random subset of Parquet files (vector column only). It runs $k$-means with $k = (\text{num\_executors}) \times (\text{partitions\_per\_executor})$ to produce a centroid codebook. This codebook becomes the IVF-style partitioning scheme that determines which vector goes to which executor's shard. The codebook is broadcast to all executors as part of the next stage's task info.

\textbf{Stage 1: Local shard build (parallel on executors).} Each executor receives an \texttt{IndexBuildTaskInfo} containing the centroid codebook, its assigned data files, and the algorithm parameters $(R, L, \alpha, \text{metric}, \text{pq\_m}, \text{pq\_nbits})$. The executor downloads the assigned Parquet files (with column projection limited to the vector column) into its local SSD cache, decodes them in streaming batches, assigns each vector to its nearest centroid to confirm shard ownership, and feeds the assigned vectors into a Vamana builder. Once built, the shard is serialized into the blob format of Section~\ref{sec:format} and written directly to an S3 object whose path was specified in the task info. The executor streams back an \texttt{IndexBuildResult} containing the shard's S3 path, vector count, byte size, and per-partition centroid distances used to populate the routing table.

\textbf{Stage 2: Assemble and commit (coordinator).} The coordinator downloads only the executor-written shard blob bytes (via S3 range request if they were uploaded as separate objects, or concatenates them into a single Puffin if uploaded inline), constructs the routing blob with the centroid codebook plus the set of covered file paths, writes the assembled Puffin file to \texttt{<table\_location>/metadata/ann-<index\_name>-snap-<snapshot\_id>.puffin}, and commits a new Iceberg snapshot via the REST catalog. The commit is structurally a REPLACE operation (no data files change) with a single new property: \texttt{summary[``statistics-file"] = puffin\_path}.

The new code introduced by this flow is small. The Vamana builder itself is the largest piece, on the order of 2{,}000 lines of C++ if implemented from scratch, or a thin wrapper if we depend on Microsoft's open-source DiskANN library~\cite{diskann-github}. The PuffinWriter is approximately 200 lines, mirroring the existing PuffinReader. The remaining components --- task dispatch, S3 I/O, cache-aware scheduling, REST commit --- are reused unchanged from the engine's INSERT-SELECT path. The total integration delta is well under 1{,}000 lines outside the algorithmic core.

\section{Distributed Probe}
\label{sec:probe}

\begin{figure}[h]
\centering
\resizebox{0.95\textwidth}{!}{%
\input{figures/tikz_probe.tex}%
}
\caption{Three-stage distributed ANN probe. Stage A runs an approximate Vamana beam search in parallel on each shard owner. The coordinator merges candidate tuples and dispatches Stage B, which reads only the row groups containing candidate vectors and computes exact distances. Stage C performs a final ordered merge.}
\label{fig:probe}
\end{figure}
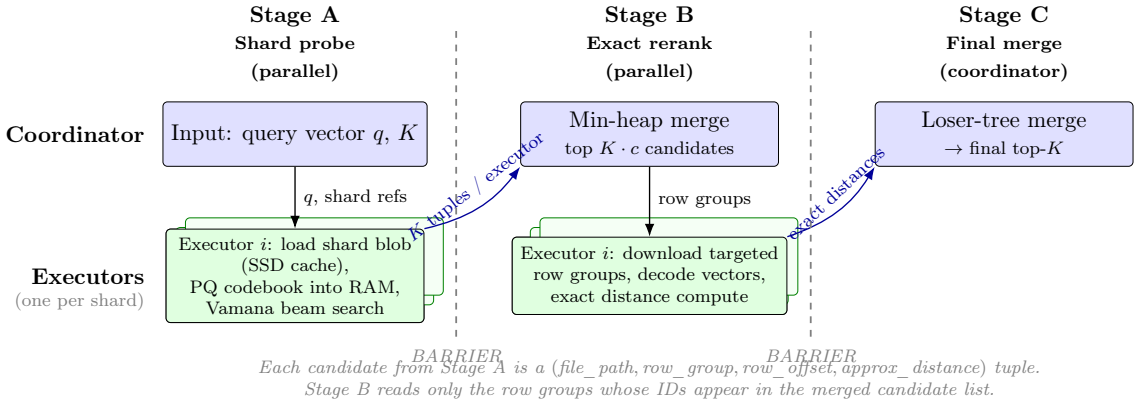

Query plans containing the pattern \texttt{ORDER BY <distance>(<column>, <literal>) LIMIT K} or \texttt{WHERE <distance>(<column>, <literal>) < threshold} are detected during planning. If an ANN index exists for the referenced column on the current snapshot, the plan is rewritten into a three-stage distributed probe, illustrated in Figure~\ref{fig:probe}.

\textbf{Stage A: Shard probe (parallel on executors).} Each executor loads its assigned shard blob from the SSD cache (or downloads it on cache miss), deserializes the Vamana graph into a memory-mapped structure, loads the PQ codebook into RAM ($\sim$4\,GB for a 250M-vector shard), and runs the standard Vamana greedy beam search with the query vector. Search proceeds with PQ-approximate distances for candidate scoring and full-precision distances for a small final rerank, traversing $\sim$500--1000 nodes per query. The executor emits the top-$K$ candidates as $(\text{file\_path}, \text{row\_group}, \text{row\_offset}, \text{approx\_distance})$ tuples via Arrow IPC.

\textbf{Stage B: Exact rerank (parallel on executors).} The coordinator merges the $K \times N$ candidate tuples from the $N$ shards using a min-heap, picks the top $K \times \text{oversample}$ candidates (oversample factor typically 2--4), maps each candidate to its source file and row-group, and dispatches per-file row-group masks to the appropriate executors. Each executor downloads only the targeted row groups, decodes the vector column for the masked rows, computes exact distances, and emits the resulting rows.

\textbf{Stage C: Final merge (coordinator).} A streaming loser-tree merge on the distance column produces the final top-$K$ result.

The pattern of (a) coarse probe producing candidates, (b) exact rerank against full-precision vectors, (c) ordered merge is structurally identical to how the engine handles \texttt{ORDER BY ... LIMIT} queries against ordinary columns, with bloom filter or zone-map pruning as the coarse stage. The ANN index slots into the same machinery; the only new component is the executor-side Vamana probe code.

\section{Incremental Refresh}
\label{sec:refresh}

A full rebuild for a billion-vector index is on the order of one hour with our parameters. We avoid it by leveraging Iceberg's manifest diff.

\subsection{Detecting What Changed}

The engine's existing \texttt{IcebergSnapshotDiffer} compares two snapshot IDs and classifies every data file as \texttt{EXISTING}, \texttt{ADDED}, or \texttt{DELETED} based on manifest entry status flags. A \texttt{REFRESH INDEX} statement reads the routing blob of the current Puffin file to discover the \texttt{base\_snapshot\_id} the index was built against, computes the diff against the current snapshot, and obtains three sets: unchanged files (whose shard contributions are still valid), added files (whose vectors must be inserted), and removed files (whose vectors must be tombstoned).

\subsection{Vamana Greedy Insert}

For each added vector $v$, the executor performs a Vamana greedy insert. The algorithm is the one described in~\cite{subramanya2019diskann}: greedy beam search from the medoid produces $L$ candidate neighbors; robust-prune selects $R$ of them using the $\alpha$-RNG criterion; forward and reverse edges are inserted; over-degree neighbors are robust-pruned back to degree $R$. The cost per insert is approximately 1{,}000 PQ-approximate distance computations plus $R$ full-precision distance computations, on the order of 200 microseconds per vector at 768 dimensions with AVX2.

For a typical incremental batch of $10^6$ new vectors distributed across 4 executors, each executor performs $\sim$250{,}000 inserts in roughly 50 seconds of compute. In practice this is dominated by the S3 read time for the new Parquet files, not the graph mutation itself.

\subsection{Lazy Tombstoning for Deletions}

Full removal of a vector from a Vamana graph is expensive: every node that has $v$ as a neighbor must be patched, and the resulting holes must be re-linked. We use lazy tombstoning instead. The shard maintains a tombstone bitmap; deleted vectors have their bit set, remain in the graph for connectivity purposes (so search traversals still function), but are filtered out of the candidate set before being returned to the coordinator. Tombstone density is recorded per shard in the routing blob:

\begin{lstlisting}
ann.shard_0.tombstone_ratio = "0.003"
ann.shard_1.tombstone_ratio = "0.15"
\end{lstlisting}

A shard whose tombstone ratio exceeds 20\% is scheduled for a full rebuild (only that shard, not the entire index) at the next maintenance window.

\subsection{Metadata-Only Commit}

The refreshed Puffin file is written to a new S3 object. The Iceberg commit is metadata-only: no new data files are added, but the snapshot summary is updated to point to the new Puffin path. The Iceberg REST specification supports this directly via the \texttt{set-properties} update action. Engines that read the snapshot will see the same data manifest as before, plus a new \texttt{statistics-file} reference; older Puffin files become unreferenced and are reaped by the table format's existing orphan-file cleanup.

\section{Implementation}
\label{sec:impl}

We implemented this design in FlockDB, an MPP analytical engine built on DuckDB~\cite{raasveldt2019duckdb} as the per-node execution kernel, with a custom distributed coordinator, an Arrow-IPC fragment protocol, and a PostgreSQL wire layer for client compatibility. The engine connects to Iceberg via a REST catalog and treats data files as immutable objects in an S3-compatible store.

The new code falls into four buckets:

\begin{itemize}
  \item \textbf{PuffinWriter} (\texttt{iceberg/puffin\_writer.cpp}): the inverse of the existing PuffinReader; produces files in the same binary layout the reader already parses. Approximately 200 lines.
  \item \textbf{SQL parsing and routing} (extensions to \texttt{query\_frontend.cpp}, \texttt{wire\_session.cpp}, \texttt{coordinator.cpp}): adds \texttt{CREATE INDEX}, \texttt{DROP INDEX}, \texttt{REFRESH INDEX} DDL handling in the \texttt{SqlLexer} fallback path, with a typed \texttt{IndexDDLInfo} struct routed to a new \texttt{IndexBuildHandler}. Approximately 500 lines.
  \item \textbf{IndexBuildHandler and planning} (\texttt{coordinator/index\_planner.cpp}, \texttt{coordinator/}\allowbreak\texttt{index\_build\_handler.cpp}): orchestrates the lifecycle, dispatches fragments through the existing scheduler, collects results, assembles the Puffin file, and commits the snapshot. Approximately 1{,}500 lines.
  \item \textbf{Vamana builder and probe} (\texttt{executor/ann\_builder.cpp}, \texttt{executor/diskann\_engine.cpp}): the algorithmic core. We currently use a from-scratch Vamana implementation rather than depending on the Microsoft DiskANN library, to keep the dependency surface small. Approximately 5{,}500 lines including PQ training, graph construction, beam search, and serialization.
\end{itemize}

The total implementation is approximately 8{,}000 lines of new code. Notably, no changes were required to the Iceberg REST catalog client beyond populating one new property in the existing commit body --- the \texttt{statistics-file} key under \texttt{summary}, which the REST endpoint already serializes verbatim as part of arbitrary summary properties.

\section{Analysis}
\label{sec:eval}

This is a design and implementation paper rather than an empirical evaluation; the system is in pre-production state at the time of writing and full benchmarks remain ongoing. We present projected resource and performance numbers for a representative configuration and the analytical reasoning behind them.

\subsection{Target Workload}

We consider a table with $10^9$ rows, each carrying a 768-dimensional float32 embedding column (approximately 300\,GB of vector data), stored as $10^4$ Parquet files in an S3-compatible object store. The engine is configured with 4 executors, each with 16 cores, 64\,GB of RAM, and 350\,GB of local NVMe SSD. Vamana parameters are $R=64$, $L=100$, $\alpha=1.2$, with PQ at $m=48$ subquantizers and 8 bits per code.

\subsection{Projected Build Cost}

A 1\% sample for centroid training is approximately 3\,GB of vector reads, completing in tens of seconds. Each executor builds a shard over approximately $2.5 \times 10^8$ vectors, with a working set dominated by the PQ-compressed representation ($2.5 \times 10^8 \times 48 = 12\,\text{GB}$ in memory) plus the graph adjacency. Empirically, Vamana builds at this scale complete in 30--60 minutes on 16-core hardware with NVMe SSD~\cite{subramanya2019diskann}; the dominant cost is the graph construction itself rather than I/O. Cross-executor coordination is limited to the broadcast of centroids in Stage 0 and the result IPC in Stage 1. Total build time is dominated by the slowest executor's shard build, projected at 45--60 minutes for a billion-vector table on 4 executors.

\subsection{Projected Query Cost}

Table~\ref{tab:query-perf} summarizes the projected per-query cost for different probe strategies on the target workload. Numbers are derived from the per-component costs in the cited literature and our own profiling of the engine's existing fan-out and merge paths.

\begin{table}[h]
\centering
\caption{Projected query performance, $10^9$ vectors, top-100 query, 4 executors.}
\label{tab:query-perf}
\small
\begin{tabular}{lrrr}
\toprule
& No index & Centroid (coord.) & DiskANN (dist.) \\
\midrule
Files scanned & 10{,}000 & $\sim$400 & $\sim$50 \\
Data read from S3 & 300\,GB & 12\,GB & 1.5\,GB \\
Cold-cache latency & 60--90\,s & 5--10\,s & 1--3\,s \\
Warm-cache latency & 15--25\,s & 2--4\,s & 50--200\,ms \\
Recall@100 (vs. exact) & 1.00 & 0.85--0.95 & 0.95--0.99 \\
\bottomrule
\end{tabular}
\end{table}

The recall trade-off for the distributed Vamana case is the main quality concern and stems directly from the independent-shard design choice (Principle~1, Section~\ref{sec:principles}). A query whose true top-$K$ neighbors are concentrated in a single shard incurs no loss; a query whose true top-$K$ spans multiple shards loses recall to the extent that the per-shard local top-$K$ misses globally-relevant candidates. Oversampling (each shard returns top-$K \times c$ candidates) and the exact rerank stage recover most of the loss. The numbers in Table~\ref{tab:query-perf} assume an oversampling factor of 4.

\subsection{Index Storage Cost}

The shard blobs occupy approximately 60\,GB each (with full-vector rerank section omitted) or 250\,GB each (with it included). The routing blob is a few megabytes. For the configuration above, the total Puffin file is on the order of 240\,GB (lean) or $\sim$1\,TB (with rerank vectors). Relative to the 300\,GB of underlying vector data, this is comparable to or somewhat larger than the data itself --- the standard trade-off of graph-based ANN indexes.

\section{Limitations}
\label{sec:limitations}

We list the limitations of the design honestly.

\textbf{Update granularity is the snapshot.} A vector cannot become searchable independent of an Iceberg commit. For streaming-insert workloads with strict freshness requirements, the snapshot commit cadence becomes the bound on index freshness. Iceberg supports streaming ingest via small frequent commits, but very high commit rates (thousands per minute) eventually saturate the REST catalog.

\textbf{Recall depends on the data distribution across shards.} The independent-shard design is statistically equivalent to IVF with one cluster per shard, when the partitioning is centroid-based. Workloads where the query distribution is poorly correlated with the data partitioning will see lower recall than workloads where it is well-correlated. We have not yet measured this systematically.

\textbf{Probe latency floor is set by the stage barriers.} The three-stage distributed probe adds two stage barriers (probe $\to$ rerank, rerank $\to$ merge), each contributing approximately 5--10\,ms in our infrastructure. For applications targeting sub-10\,ms P50 latency this is too much; a single-stage local index would be more appropriate.

\textbf{Object store throughput bounds cold-cache performance.} On a cold cache, the probe and rerank phases must read shard blobs and Parquet row groups from S3 respectively. Cold-cache P99 is bounded by S3 latency and throughput, not by the engine.

\textbf{No formal recall guarantee.} The design provides no analytical lower bound on recall. Existing literature on Vamana provides empirical recall numbers for the algorithm itself, but the independent-shard composition is not analyzed there.

\textbf{Concurrent build/refresh resolution relies on Iceberg optimistic concurrency.} Two coordinators attempting to refresh the same index simultaneously will see a commit conflict at the REST layer; one will succeed, the other must retry against the new base snapshot. We do not currently coalesce or queue concurrent refreshes; an operator scheduling background refreshes alongside user-issued refreshes can produce wasted work.

\section{Discussion and Future Work}

The pattern we describe is not specific to ANN. Anything that needs to be derived from table data, versioned in lockstep with snapshots, and read by some engines while ignored by others, is a candidate for Puffin storage. We see at least three further applications worth exploring:

\textbf{Learned indexes and cardinality models.} Recurrent neural networks and learned distribution models trained per column~\cite{kraska2018case} can be stored as Puffin blobs and consumed by query optimizers that know how to use them.

\textbf{Materialized derived columns.} Computed aggregates, derived embeddings, or denormalized join columns can be stored as Puffin blobs and queried as virtual columns, with consistency to the underlying data guaranteed by snapshot binding.

\textbf{Cross-engine acceleration sharing.} An ANN index built by one engine could in principle be consumed by another, provided they agree on the blob format. We propose a family of three Puffin blob types --- \texttt{ann-routing-v1}, \texttt{ann-vamana-graph-v1}, and \texttt{ann-centroid-index-v1} --- as candidate community standards, with the layouts described in Section~\ref{sec:format}. A separate companion document defining the wire format at the level of detail required for a Puffin specification revision is available alongside this paper. Whether engines will converge on a standard is a community question; the Iceberg \texttt{dev@} list is the appropriate forum.

The broader observation is that the open table format provides a derived-data lifecycle for free. We expect more engines to begin using Puffin not merely for table statistics but as a general-purpose append-only key-value store bound to snapshot semantics. The pattern transfers cleanly to Delta Lake's deletion vectors and statistics files~\cite{armbrust2020delta}, which occupy similar architectural slots, and to Hudi's metadata table~\cite{hudi}.

\section{Conclusion}

We have described a design and implementation for placing distributed ANN indexes inside Apache Iceberg Puffin files, with the index lifecycle bound to the table format's snapshot lifecycle. The pattern preserves the compute-disaggregated invariants of a modern lakehouse engine: executors remain stateless, derived state is durably stored in object storage and snapshot-versioned, and no new lifecycle is required for index creation, refresh, or deletion. The recall trade-off introduced by the independent-shard design is real but well-understood and bounded by oversampling. We believe the pattern is the most natural way to add vector search to lakehouse-style engines, and we have begun upstreaming the relevant blob-type definitions for community discussion.

\bibliographystyle{plain}
\bibliography{references}

\end{document}

%% file: figures/tikz_overview.tex
\begin{tikzpicture}[
    font=\small,
    box/.style={draw, rounded corners=2pt, align=center, inner sep=4pt},
    coord/.style={box, fill=blue!12, minimum width=110mm, minimum height=14mm},
    exec/.style={box, fill=green!12, minimum width=26mm, minimum height=18mm},
    catalog/.style={box, fill=orange!18, minimum width=30mm, minimum height=12mm},
    s3/.style={
      draw, cylinder, shape border rotate=90, aspect=0.18,
      minimum width=42mm, minimum height=22mm, fill=gray!12,
      text width=32mm, align=center
    },
    client/.style={box, fill=yellow!25, minimum width=26mm, minimum height=10mm},
    flow/.style={-{Latex[length=2.2mm]}, semithick},
    dashflow/.style={-{Latex[length=2.2mm]}, semithick, dashed},
    note/.style={font=\scriptsize\itshape, gray}
  ]

  \node[client] (client) {Client \scriptsize(psql, JDBC)};

  \node[coord, below=6mm of client] (coord)
    {\textbf{Coordinator}\\[1pt]
     \scriptsize QueryFrontend $\,\cdot\,$ DistributedPlanner $\,\cdot\,$ IndexBuildHandler $\,\cdot\,$ PuffinWriter};

  \node[exec, below=14mm of coord.south, xshift=-42mm] (e0)
    {\textbf{Executor 0}\\[1pt]\scriptsize DuckDB kernel\\\scriptsize AnnBuilder\\\scriptsize SSD L2 cache};
  \node[exec, right=4mm of e0] (e1)
    {\textbf{Executor 1}\\[1pt]\scriptsize DuckDB kernel\\\scriptsize AnnBuilder\\\scriptsize SSD L2 cache};
  \node[exec, right=4mm of e1] (e2)
    {\textbf{Executor 2}\\[1pt]\scriptsize DuckDB kernel\\\scriptsize AnnBuilder\\\scriptsize SSD L2 cache};
  \node[exec, right=4mm of e2] (e3)
    {\textbf{Executor 3}\\[1pt]\scriptsize DuckDB kernel\\\scriptsize AnnBuilder\\\scriptsize SSD L2 cache};

  \node[s3, below=20mm of e1.south, xshift=12mm] (s3)
    {\textbf{S3} / object store \\\scriptsize Parquet $\,\cdot\,$ \texttt{*.puffin}};
  \node[catalog, left=22mm of s3] (catalog) {Iceberg\\REST Catalog};

  \draw[flow] (client) -- node[right, font=\scriptsize] {SQL} (coord);

  \draw[flow] (coord.south) -- node[left, font=\scriptsize, pos=0.65] {fragments} (e0.north);
  \draw[flow] (coord.south) -- (e1.north);
  \draw[flow] (coord.south) -- (e2.north);
  \draw[flow] (coord.south) -- node[right, font=\scriptsize, pos=0.65] {Arrow IPC} (e3.north);

  \draw[flow] (coord.west) to[out=180,in=180,looseness=1.3]
        node[left, font=\scriptsize, pos=0.5] {catalog API} (catalog.west);

  \draw[dashflow] (catalog.east) -- node[above, font=\scriptsize] {\texttt{statistics-file}} (s3.west);

  \node[fit=(e0)(e1)(e2)(e3), inner sep=0pt] (execgroup) {};
  \draw[flow] (e0.south) -- ++(0,-2mm) -| (s3.north);
  \draw[flow] (e1.south) -- ++(0,-4mm) -| ([xshift=-6mm]s3.north);
  \draw[flow] (e2.south) -- ++(0,-4mm) -| ([xshift=6mm]s3.north);
  \draw[flow] (e3.south) -- ++(0,-2mm) -| (s3.north);

  \node[font=\scriptsize, gray] at ($(execgroup.south)+(0,-9mm)$) {Parquet reads $\,\cdot\,$ Puffin shard writes};

  \node[note, above=0.5mm of client.north east, anchor=south west, xshift=4mm] {Compute layer (stateless)};
  \node[note, below=2mm of s3.south, anchor=north] {Storage layer (durable, object store + catalog)};

\end{tikzpicture}

%% file: figures/tikz_puffin_layout.tex
\begin{tikzpicture}[
    font=\small,
    block/.style={draw, rectangle, minimum width=72mm, align=center, inner sep=4pt},
    magic/.style={block, fill=gray!20, minimum height=6mm, font=\ttfamily\small},
    blob/.style={block, fill=blue!10, minimum height=18mm},
    routingblob/.style={block, fill=orange!18, minimum height=14mm},
    footer/.style={block, fill=green!12, minimum height=12mm},
    tail/.style={block, fill=gray!12, minimum height=6mm, font=\scriptsize\ttfamily},
    snap/.style={draw, rounded corners=2pt, fill=yellow!22, minimum width=46mm, minimum height=12mm, align=center, font=\small},
    flow/.style={-{Latex[length=2mm]}, semithick},
    offset/.style={font=\scriptsize\ttfamily, gray},
    annot/.style={font=\scriptsize, align=left}
  ]

  \node[snap] (snap) {Iceberg snapshot $S_n$\\[1pt]\scriptsize\ttfamily summary["statistics-file"]};

  \node[magic, right=22mm of snap, yshift=42mm] (m1) {PFA1 \footnotesize(4B magic)};
  \node[routingblob, below=0pt of m1] (routing) {%
     \textbf{Blob 0:} \texttt{flockdb-ann-routing-v1}\\[1pt]
     \scriptsize fields=[7]\ \ algorithm=diskann\\
     \scriptsize centroids, shard map, covered files};
  \node[blob, below=0pt of routing] (b1) {%
     \textbf{Blob 1:} \texttt{flockdb-ann-index-v1} \scriptsize(shard 0)\\[1pt]
     \scriptsize PQ codebook $\,\cdot\,$ adjacency $\,\cdot\,$ rerank vectors\\
     \scriptsize vector\_id $\to$ (file, row\_group, offset)};
  \node[blob, below=0pt of b1] (b2) {%
     \textbf{Blob 2:} \texttt{flockdb-ann-index-v1} \scriptsize(shard 1)\\[1pt]
     \scriptsize $\sim$60--250 GB per shard};
  \node[blob, below=0pt of b2, minimum height=12mm] (bdots) {$\cdots$ shards 2 $\ldots$ $N-1$ $\cdots$};
  \node[footer, below=0pt of bdots] (foot) {%
     \textbf{Footer:} UTF-8 JSON\\[1pt]
     \scriptsize \texttt{\{"blobs":[\{type, fields, offset, length, compression, props\}, $\ldots$]\}}};
  \node[tail, below=0pt of foot] (t1) {footer\_size (4B LE) $\,\cdot\,$ flags (4B) $\,\cdot\,$ PFA1 magic (4B)};

  \node[draw, thick, rounded corners=3pt, gray,
        fit=(m1)(routing)(b1)(b2)(bdots)(foot)(t1),
        inner sep=2pt] (puffin) {};
  \node[above=0pt of puffin.north, font=\small\bfseries] {\texttt{ann-embedding\_idx-snap-$S_n$.puffin}};

  \draw[flow] (snap.east) -- ++(8mm,0) |- (puffin.west);

  \node[offset, right=3mm of m1.east, anchor=west] {offset 0};
  \node[offset, right=3mm of routing.east, anchor=west] {offset 4};
  \node[offset, right=3mm of b1.east, anchor=west] {offset 4+|B0|};
  \node[offset, right=3mm of foot.east, anchor=west] {file\_size $-$ 12 $-$ |F|};
  \node[offset, right=3mm of t1.east, anchor=west] {file\_size $-$ 12};

  \node[annot, left=4mm of routing.west, anchor=east, text width=20mm] {small\\ (few MB)\\ probe-time\\ read};
  \node[annot, left=4mm of b1.west, anchor=east, text width=20mm] {large\\ executor-owned\\ range-read};

\end{tikzpicture}

%% file: figures/tikz_probe.tex
\begin{tikzpicture}[
    font=\small,
    coordbox/.style={draw, rounded corners=2pt, fill=blue!12,
                     minimum width=40mm, minimum height=10mm, align=center},
    execbox/.style={draw, rounded corners=2pt, fill=green!12,
                    minimum width=40mm, minimum height=12mm, align=center, font=\scriptsize},
    barrier/.style={draw, dashed, thick, gray},
    flow/.style={-{Latex[length=2mm]}, semithick},
    dataflow/.style={-{Latex[length=2mm]}, semithick, blue!60!black},
    stagehdr/.style={font=\small\bfseries, align=center},
    note/.style={font=\scriptsize\itshape, gray, align=center}
  ]

  \def\xA{0}
  \def\xB{55}
  \def\xC{110}

  \node[stagehdr] (hA) at (\xA mm, 30mm) {Stage A\\\scriptsize Shard probe\\\scriptsize (parallel)};
  \node[stagehdr] (hB) at (\xB mm, 30mm) {Stage B\\\scriptsize Exact rerank\\\scriptsize (parallel)};
  \node[stagehdr] (hC) at (\xC mm, 30mm) {Stage C\\\scriptsize Final merge\\\scriptsize (coordinator)};

  \node[coordbox] (cInput) at (\xA mm, 16mm)
    {Input: query vector $q$, $K$};
  \node[coordbox] (cMerge) at (\xB mm, 16mm)
    {Min-heap merge\\\scriptsize top $K \cdot c$ candidates};
  \node[coordbox] (cFinal) at (\xC mm, 16mm)
    {Loser-tree merge\\\scriptsize $\to$ final top-$K$};

  \node[execbox] (eA) at (\xA mm, -6mm)
    {Executor $i$: load shard blob\\(SSD cache),\\PQ codebook into RAM,\\Vamana beam search};
  \node[execbox] (eB) at (\xB mm, -6mm)
    {Executor $i$: download targeted\\row groups, decode vectors,\\exact distance compute};

  \node[execbox, draw=green!50!black, fill=green!4,
        minimum width=40mm, minimum height=12mm] at ($(eA) + (1.5mm, 1.5mm)$) {};
  \node[execbox, draw=green!50!black, fill=green!4,
        minimum width=40mm, minimum height=12mm] at ($(eA) + (3mm, 3mm)$) {};
  \node[execbox] (eAfront) at (\xA mm, -6mm)
    {Executor $i$: load shard blob\\(SSD cache),\\PQ codebook into RAM,\\Vamana beam search};
  \node[execbox, draw=green!50!black, fill=green!4,
        minimum width=40mm, minimum height=12mm] at ($(eB) + (1.5mm, 1.5mm)$) {};
  \node[execbox, draw=green!50!black, fill=green!4,
        minimum width=40mm, minimum height=12mm] at ($(eB) + (3mm, 3mm)$) {};
  \node[execbox] (eBfront) at (\xB mm, -6mm)
    {Executor $i$: download targeted\\row groups, decode vectors,\\exact distance compute};

  \draw[flow] (cInput.south) -- node[right, font=\scriptsize] {$q$, shard refs} (eAfront.north);
  \draw[dataflow] (eAfront.north east) to[bend right=20]
        node[above, font=\scriptsize, sloped, pos=0.6] {$K$ tuples / executor} (cMerge.south west);

  \draw[flow] (cMerge.south) -- node[right, font=\scriptsize] {row groups} (eBfront.north);
  \draw[dataflow] (eBfront.north east) to[bend right=20]
        node[above, font=\scriptsize, sloped, pos=0.6] {exact distances} (cFinal.south west);

  \draw[barrier] (\xA mm + 25mm, 32mm) -- (\xA mm + 25mm, -16mm)
        node[below, font=\scriptsize\itshape, gray] {BARRIER};
  \draw[barrier] (\xB mm + 25mm, 32mm) -- (\xB mm + 25mm, -16mm)
        node[below, font=\scriptsize\itshape, gray] {BARRIER};

  \node[font=\small\bfseries, anchor=east] at (-22mm, 16mm) {Coordinator};
  \node[font=\small\bfseries, anchor=east] at (-22mm, -6mm) {Executors};
  \node[font=\scriptsize, gray, anchor=east] at (-22mm, -10mm) {(one per shard)};

  \node[note] at (\xB mm, -22mm)
    {Each candidate from Stage A is a $(\text{file\_path}, \text{row\_group}, \text{row\_offset}, \text{approx\_distance})$ tuple.\\
     Stage B reads only the row groups whose IDs appear in the merged candidate list.};

\end{tikzpicture}